\documentclass[twocolumn,amsmath,ammsymb]{revtex4-1}
\usepackage{ulem}
\usepackage{epsfig}
\usepackage{amsmath}
\usepackage{amssymb}
\usepackage{graphicx}
\usepackage{dcolumn}
\usepackage{soul}
\usepackage{bm}
\usepackage[usenames]{color}
\usepackage[breaklinks]{hyperref}

\begin{document} 

\preprint{}

\title{Disorder and magnetoconductivity in tilted Weyl semimetals}
\author{Yi-Xiang Wang$^1$ and Fuxiang Li$^2$}
\affiliation{$^1$School of Science, Jiangnan University, Wuxi 214122, China.}
\affiliation{$^2$School of Physics and Electronics, Hunan University, Changsha 410082, China}
\date{\today}

\begin{abstract}
Gapless Weyl semimetals (WSMs) are a novel class of topological materials that host massless Weyl fermions as their low-energy excitations.  When the Weyl cone is tilted, the Lorentz invariance is broken and the Lifshitz transition drives the system from type-I WSMs into type-II WSMs.  Here, based on the lattice model, we perform a systematic numerical study of the effect of onsite disorder on the diagonal magnetoconductivity in tilted WSMs.  We use the self-consistent Born approximation to calculate the disorder-induced self-energy and then apply the Kubo-Streda formula to calculate the diagonal magnetoconductivity.  For the transverse magnetoconductivity $\sigma_{xx}$, the disorder is found to show distinct effects in the quantum limit regime and in the quantum oscillation regime of type-I WSMs, which could be understood from the Einstein's relationship.  With strong disorder, $\sigma_{xx}$ shows a linear relation with the inverse magnetic field $\frac{1}{B}$, which exhibits certain robustness to both the Fermi energy and the Weyl cone tilting.  For the longitudinal magnetoconductivity $\sigma_{zz}$, the strong disorder can break the positive magnetoconductivity as well as the Shubnikov-de Haas oscillations.  By analyzing the spectral function, we find that the chiral anomaly is still preserved at strong disorder even when the Weyl cone is overtilted, as there is no gap opening around the Weyl nodes.  The implications of our results for experiments are discussed. 
\end{abstract}

\maketitle

\section{Introduction}

The topological phases of matter are of immense interests due to their fundamental physics as well as the potential applications~\cite{A.Bansil, N.P.Armitage, X.Wan}.  Among the topological phases, the three-dimensional (3D) Weyl semimetals (WSMs), which host the linear dispersing quasiparticles with distinct chiralities, have been discovered in experiment and lie at the forefront of the modern condensed matter physics~\cite{S.Y.Xu2015a, S.M.Huang, B.Q.Lv, S.Y.Xu2015b}.  The 3D WSMs exhibit a lot of exotic phenomena that are not present in the traditional systems.  Among them, one interesting feature is the chiral anomaly.  In WSMs, the number of Weyl fermions with opposite chiralities is separately conserved in the absence of any gauge field coupling.  However, in the presence of the nonorthogonal electric field and magnetic field, the Weyl fermions can be pumped from one node to the other with opposite chirality, leading to the violation of separate number conservation laws and thus the emergence of the chiral anomaly.  In experiment, the observation of the negative magnetoresistivity (or positive magnetoconductivity) was believed to be a signature of the chiral anomaly~\cite{Q.Li, X.Huang, J.Xiong, C.L.Zhang, H.Li}.  Theoretical studies based on the semiclassical Boltzmann transport theory~\cite{D.T.Son} also supported this idea.  However, whether the conclusion is reliable is still under heated debates~\cite{E.V.Gorbar, H.Z.Lu, H.P.Sun, A.A.Burkov2015, H.W.Wang}.  For example, one debate considers that the negative magnetoresistivity is strongly related to the current jetting effects~\cite{N.P.Armitage}, \textit{i.e.}, the current becomes narrowly directly along the applied field, but not to the chiral anomaly.  Moreover, it was reported recently that, in the Dirac semimetal Cd$_3$As$_2$~\cite{T.Liang, Y.Zhao, J.Feng}, the magnetoresistance shows a linear and nonsaturated behavior when the magnetic field is perpendicular to the electric field.

The novel type-II WSMs were theoretically proposed~\cite{A.A.Soluyanov, Z.Wang} and soon demonstrated in the crystals of MoTe$_2$~\cite{K.Deng, L.Huang, A.Tamai}, WTe$_2$~\cite{C.Wang} and the alloy of Mo$_x$W$_{1-x}$Te$_2$~\cite{I.Belopolski2016a, I.Belopolski2016b}.  Unlike the type-I WSMs, the linear band dispersions around a Weyl node in type-II WSMs are significantly tilted.  Consequently, the Fermi surface encloses both the electron and hole pockets, and the density of states (DOS) can be nonzero even at the Weyl node.  Thus for the overtilted Weyl nodes, the unconventional Fermi surface can mask the contributions from the Weyl nodes.  A nontrivial question is how to detect the transition from type-I WSMs to type-II and to recover the concealed Weyl nodes in experiment.  The magnetotransport measurement may provide a possible avenue to investigate this transition.  Indeed, the previous magnetic-optical response studies revealed that the anomalous resonance absorption peaks were closely connected to the chiral zeroth Landau Levels (LLs) in type-II WSMs~\cite{Z.M.Yu, M.Udagawa, S.Tchoumakov}.

In theory, Abrikosov~\cite{A.A.Abrikosov} initially analyzed the magnetoconductivity in a Dirac semimetal.  He considered  the long-range charged impurities scattering,  and within the Born approximation, he obtained a linear magnetoresistance when the chemical potential coincides with the zeroth LL.  Recently the magnetotransport problem  have been reexamined in WSMs by many researchers~\cite{E.V.Gorbar, J.Klier2015, J.Klier2017, X.Xiao, H.W.Wang}.  In particular, different models of short-range impurities and charged (Coulomb) impurities have been proposed for the study of transverse magnetoconductivity~\cite{J.Klier2015,J.Klier2017}, with a rich variety of  conductivity scaling regimes being identified.  Xiao \textit{et.al}~\cite{X.Xiao} analyzed the effect of the chemical potential and temperature on the magnetoconductivity in WSMs. 
 
There are also several theoretical works about the magnetotransport in tilted WSMs~\cite{V.A.Zyuzin, K.Das, Y.W.Wei, G.Sharma, Y.X.Wang2019}.  Within the low-energy approximation and by using the semiclassical Boltzmann transport theory~\cite{V.A.Zyuzin, K.Das, G.Sharma}, it was revealed that the chiral-anomaly-induced positive longitudinal magnetoconductivity is still present in type-II WSMs.  However, a question arises that when the Weyl cone is overtilted, the high-energy states, together with the low-energy ones, will participate in the magnetotransport.  So the low-energy model may not correctly capture the closed Fermi pockets.  Therefore, to more accurately analyze the magnetotransport in type-II WSMs, the tight-binding lattice model is needed to access the full extent of the band tilting in the momentum space.  Another motivation is that we try to study the effect of onsite disorder on the magnetotransport in tilted WSMs.  To our knowledge, there are few works discussing this problem~\cite{Y.W.Wei, Y.X.Wang2019}.  In the work by Wei and \textit{et al.}~\cite{Y.W.Wei}, the longitudinal magnetoconductivity in type-II WSMs was considered, with emphasis on the effect of the range of the impurity potentials.  In our previous work~\cite{Y.X.Wang2019}, by using the exact diagonalization, the effect of disorder on the Hall conductivity was studied in tilted WSMs and several striking signatures were found to distinguish type-I WSMs from type-II.  

Motivated by these theoretical and experimental progresses, in this paper, we investigate the effect of disorder on the diagonal magnetoconductivity in tilted WSMs.  Based on the minimal lattice model, the self-consistent Born approximation is used to calculate the self-energy induced by the onsite disorder.  Then we apply the Kubo-Streda formula to calculate the transverse and longitudinal magnetoconductivity.  Considering that experimently the Fermi energy in WSMs is usually away from zero due to the finite carrier density,  it is crucial to understand the roles played not only by the zeroth LL, but also by the $n\ge1$ LLs in the magnetotransport.  We make a systematic study of how the diagonal magnetoconductivity is affected by the onsite disorder, Fermi energy, magnetic field and Weyl cone tilting.   

Our main results are as follows. For the transverse magnetoconductivity $\sigma_{xx}$, disorder can drive the crossover of type-I WSMs from the 3D Hall state into the diffusive metal state.  More importantly, the effect of disorder is shown to be distinct in the quantum limit regime and quantum oscillation regime of type-I WSMs, which could be understood by the Einstein's relationship.  At strong disorder, the linear relationship of $\sigma_{xx}$ with the inverse magnetic field $\frac{1}{B}$ shows certain robustness, which is not limited in the quantum limit regime of nontilted WSMs, but also is observed in both type-I and type-II WSMs for varying Fermi energy.  So $\sigma_{xx}$ cannot be saturated with the magnetic field.  However, it is interesting to find that strong disorder can drive $\sigma_{xx}$ to reach its saturation value even for the overtilted Weyl cone.   For the longitudinal magnetoconductivity $\sigma_{zz}$, the strong disorder can break the positive magnetoconductivity as well as the Shubnikov-de Haas (SdH) oscillations.  By analyzing the spectral function, we find that the chiral anomaly is still preserved at strong disorder even when the Weyl cone is overtilted, as there is no gap opening around the Weyl nodes.  In addition, the SdH oscillation frequency in $\sigma_{zz}$ is studied and exhibits discrete steps with the Weyl cone tilting.  The different behaviors of $\sigma_{xx}$ and $\sigma_{zz}$ reflect the Weyl cone anisotropy when both the magnetic field and Weyl cone tilting are present.  The obtained results may deepen the understandings about the interplay between the magnetic field and disorder in tilted WSMs.

\section{Model and Method}

We start from the model that hosts a pair of Weyl nodes.  The model is constructed on a simple cubic lattice with two orbitals on each site.  The lattice constant is set as $a_0=1$.  Without the Weyl cone tilting, the model is assumed to break the time-reversal symmetry, but preserve the inversion symmetry.  This allows for the minimal number of Weyl nodes, two with opposite chiralities.  Including the tilting term, the continuous Hamiltonian reads~\cite{T.M.McCormick, Y.X.Wang2017, Y.X.Wang2019, M.J.Park, Y.Wu}: 
\begin{align}
H_0(\boldsymbol k)
=&2t(\text{cos}k_x\sigma_x+\text{sin}k_y\sigma_y+\text{sin}k_z\sigma_z)
\nonumber\\
&+m_0(2-\text{cos}k_y-\text{cos}k_z)\sigma_x+2t_z\text{sin}k_z\sigma_0.
\end{align}
Here the Pauli matrice $\sigma$ acts on the orbital and the Wilson mass term $m_0$ is used to open a finite energy window as to avoid the band overlapping.  The two Weyl nodes are located at $\boldsymbol K_\eta=(\eta\frac{\pi}{2},0,0)$, with $\eta=\pm$.  When $t_z=0$, the inversion symmetry is given as $H(\boldsymbol k)=\sigma_xH(-\boldsymbol k)\sigma_x$.  The introduction of $t_z$  makes the Weyl cone anisotropic and breaks the inversion symmetry. Here we choose the Weyl cone tilting in $k_z-$direction, perpendicular to the distance between the Weyl nodes.  Such a choice of $t_z$ can be easily tuned to feature the electron and hole pockets in type-II WSMs~\cite{M.Trescher}.  Around the Weyl node $\boldsymbol K_\eta$, the Hamiltonian $H_0$ is expanded to yield a low-energy continuous description, 
\begin{align}
H_{0\eta}(\boldsymbol k)=\hbar v(-\eta k_x\sigma_x+k_y\sigma_y+k_z\sigma_z)+\hbar v_zk_z\sigma_0,
\end{align}
with the velocities $v=\frac{2t}{\hbar}$ and $v_z=\frac{2t_z}{\hbar}$.  In the following, we will use the hopping integral $t$ as the unit of energy and set $m_0=2$.

Consider the magnetic field and the quantized LLs.  If the magnetic field acts in the $x-y$ plane, the chiral zeroth modes become indistinguishable from other LLs~\cite{M.Udagawa}.  To better analyze the magnetoconductivity based on the LLs, we assume the magnetic field to be along the $z-$direction, $\boldsymbol B=(0,0,B)$.  The effect of the magnetic field is included in the system by using the Peierls substitution, $\boldsymbol p\rightarrow \boldsymbol p-e\boldsymbol A$, with the vector potential being taken in the Landau gauge of $\boldsymbol A(\boldsymbol r)=(-yB,0,0)$.  In the real space, we take the lattice size as $L_x\times L_y\times L_z$, and impose periodic boundary conditions for all three directions.  The Hamiltonian under the magnetic field is discretized as~\cite{M.Udagawa}
\begin{align}
H_0=&\sum_{\boldsymbol r,s,s'}
\{ [e^{-i yB}c^\dagger_{\boldsymbol r+x,s}t\sigma^x_{ss'} 
+c^\dagger_{\boldsymbol r+z,s} (-\frac{m_0}{2}\sigma^x_{ss'}-it\sigma^z_{ss'}
\nonumber\\
&-it_z\sigma^0_{ss'})+c^\dagger_{\boldsymbol r+y,s}(-it \sigma^y_{ss'}
-\frac{m}{2}\sigma^x_{ss'})]
c_{\boldsymbol r,s'}+\text{H.c.}\}
\nonumber\\
&+\sum_{\boldsymbol r,s,s'}
c^\dagger_{\boldsymbol r,s}2m_0\sigma^x_{ss'}c_{\boldsymbol r,s'}. 
\end{align}
Here $\boldsymbol r=(x,y,z)$ is the coordinate on a cubic lattice.  To be commensurate with the lattice structure, a common way is to write the magnetic field as $B=\frac{2\pi}{L_y}$~\cite{M.Udagawa, Y.W.Wei, Y.X.Wang2019}, with the unit of $\frac{\hbar}{ea_0^2}$. 

When $t_z=0$, the Weyl cone is nontilted and the energy window spans the region of $[-2t,2t]$.  When $t_z\neq0$, the Weyl cone, together with the energy window is tilted.  Note that we still denote the LLs inside the energy window as the low-energy LLs and those outside the energy window as the high-energy LLs~\cite{Y.X.Wang2019}.  The energies of the low-energy LLs are obtained directly by using the ladder operators, 
\begin{align}
&\varepsilon_{n\neq 0}(k_z)=\text{sgn}(n)
\sqrt{\hbar^2v^2k_z^2+\frac{2\hbar^2v^2|n|}{l_B^2}}+\hbar v_zk_z,
\\
&\varepsilon_{n=0,\eta}(k_z)=\hbar(-\eta v+v_z)k_z, 
\end{align}
with the magnetic length $l_B=\sqrt{\frac{\hbar}{eB}}$.  From Eq.~(5), it shows that the Weyl cone tilting can drastically change the properties of the chiral zeroth LLs.  When $t_z<t$, the system lies in type-I WSM and the zeroth LLs own opposite velocities and are counterpropagating.  When $t_z>t$, the lifshitz transition happens, driving the system to be  type-II WSM, in which the two zeroth LLs acquire the velocities in the same direction.  

The disordered impurity potential plays an indispensable role for the diagonal magnetoconductivity.  To include disorder, we consider the following form
\begin{align}
H_d=\sum_{\boldsymbol r,s} \epsilon_{\boldsymbol r,s}
\hat c^\dagger_{\boldsymbol r,s} \hat c_{\boldsymbol r,s}, 
\end{align}
with $\epsilon_{\boldsymbol r,s}$ denoting the impurity potential produced at site $\boldsymbol r$ and orbital $s$.  We assume $\epsilon_{\boldsymbol r,s}$ to be uniformly distributed in the range of $[-\frac{W}{2},\frac{W}{2}]$, with $W$ being the disorder strength.  Such a model is widely used to mimic the random on-site disorder potential in the Dirac/Weyl semimetal systems~\cite{Y.W.Wei, M.J.Park, C.Z.Chen, S.Bera, J.H.Pixley, Y.Wu}.  As the disorder configuration does not preserve the time-reversal symmetry, both the charge and magnetic disorder are included.  For a given disorder, its average value is zero, $\langle\epsilon_{\boldsymbol r,s}\rangle=0$. 

The disorder-averaged Green's function of the system is defined as 
\begin{align}
\bar G^R(\varepsilon)
=\langle\frac{1}{\varepsilon-(H_0+H_d)+i0^+}\rangle
=\frac{1}{\varepsilon-H_0-\Sigma+i0^+}. 
\end{align}
Here the self-energy $\Sigma$ is introduced to represent the effect of disorder.  Within the framework of the self-consistent Born approximation, the self-energy is calculated as~\cite{C.W.Groth, A.Altland, C.Z.Chen, Y.W.Wei}
\begin{align}
\Sigma(\varepsilon)
&=\frac{\int_{-\frac{W}{2}}^{\frac{W}{2}}\epsilon^2d\epsilon}{W}
\langle r_i|\bar G^R(\varepsilon)|r_i\rangle
\nonumber\\
&=\frac{W^2}{12}\int_{\text{MBZ}}\frac{d \boldsymbol k}{(2\pi)^3}
\frac{1}{\varepsilon-H_{\boldsymbol k}-\Sigma+i0^+}, 
\label{Self-energy}
\end{align} 
with the integration being performed over the magnetic Brillouin zone (MBZ).  Although the translational symmetry is broken by the magnetic field in the real space, it will get restored by the Peierls substitution within the magnetic supercell.  In Eq.~(\ref{Self-energy}), the self-energy is independent of the momentum and is given as a function of the energy.  The imaginary part of the self-energy is directly related to the inverse relaxation time by $\frac{1}{\tau_n}=\frac{2}{\hbar}|\text{Im}\Sigma_{nn}|$.  In the Born approximation, $\Sigma$ is solved directly and without self-consistency, by neglecting $\Sigma$ in the right-hand side of Eq.~(\ref{Self-energy})~\cite{C.W.Groth, M.J.Park}.  Here we take the standard iterative steps to calculate $\Sigma$ self-consistently and set the convergence precision between two consecutive steps to be $10^{-6}$. 

After obtaining the self-energy $\Sigma$ and the disorder-averaged Green's function $\bar G^R(\varepsilon)$, we use the Kubo-Streda formula to calculate the diagonal conductivity in the disordered system~\cite{A.Bastin,P.Streda}, 
\begin{align}
\sigma_{\alpha\alpha}(\varepsilon)
=\frac{e^2\hbar}{\pi V}\sum_{\boldsymbol k}\text{Tr}
[\hat v_\alpha\text{Im}\bar G^R(\varepsilon)\hat v_\alpha\text{Im}\bar G^R(\varepsilon)], 
\end{align}
here $V$ is the volume of the magnetic unit cell and $\hat v_\alpha=\frac{1}{i\hbar}[\hat r_\alpha,H_0]$ is the velocity operator along $\alpha$ direction.  We will focus on the transverse magnetoconductivity $\sigma_{xx}$ and the longitudinal magnetoconductivity $\sigma_{zz}$.  In experiment, $\sigma_{xx}$ is related to the configuration of the perpendicular electric field and magnetic field, $\boldsymbol E\perp\boldsymbol B$, while $\sigma_{zz}$ refers to the configuration of the parallel electric field and magnetic field, $\boldsymbol E//\boldsymbol B$, where the chiral anomaly appears.  In this work, we consider the zero temperature case.

\section{Transverse magnetoconductivity}

\subsection{Fermi Energy}

First of all, we investigate the variation of the transverse magnetoconductivity $\sigma_{xx}$ with the Fermi energy $\varepsilon$, as the Fermi energy is usually away from zero due to the finite carrier density in real samples.  For weakly tilted type-I WSMs, two regimes need to be distinguished~\cite{H.W.Wang, X.Xiao}: the quantum limit regime and the quantum oscillation regime.  The former refers to the regime of $\varepsilon<\varepsilon_{1v}$, with
\begin{align}
\varepsilon_{nv}=\frac{2}{l_B}\sqrt{2n(t^2-t_z^2)}
\end{align}
being the vertex energy of the dispersive $n-$LL.  In this regime,  the Fermi energy is located on the lowest zeroth LL and thus all charge carriers are confined on the zeroth LL.  The latter refers to the regime of $\varepsilon_{1v}<\varepsilon<2(t-t_z)$ so that the Fermi energy is located on the $n\ge1$ LL.  In this regime, besides the zeroth LL, the $n\ge1$ LLs also contribute to the magnetotransport. 

\begin{figure}
	\includegraphics[width=8.2cm]{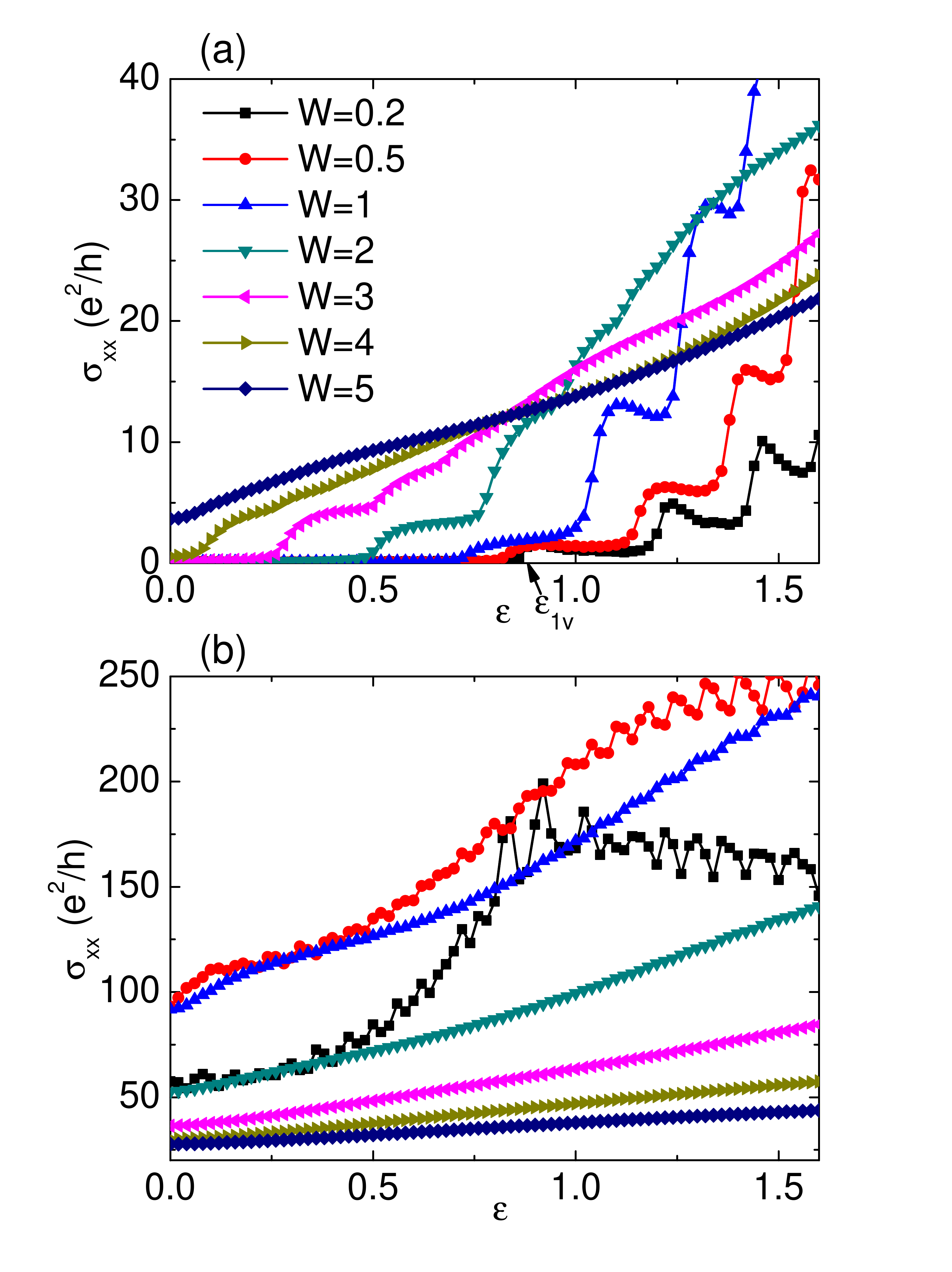}
	\caption{(Color online) The transverse magnetoconductivity $\sigma_{xx}$ vs the Fermi energy $\varepsilon$ for different disorder strength $W$.  (a) is for the nontilted WSM of $t_z=0$ and (b) is for the overtilted WSM of $t_z=1.5$.  In (a), $\varepsilon_{1v}$ is indicated by the arrow, to distinguish the quantum limit regime and quantum oscillation regime.  The legends are the same in both figures and the magnetic field is set as $B=\frac{2\pi}{60}$. }
	\label{Fig1}
\end{figure} 

\begin{figure}
	\includegraphics[width=8.2cm]{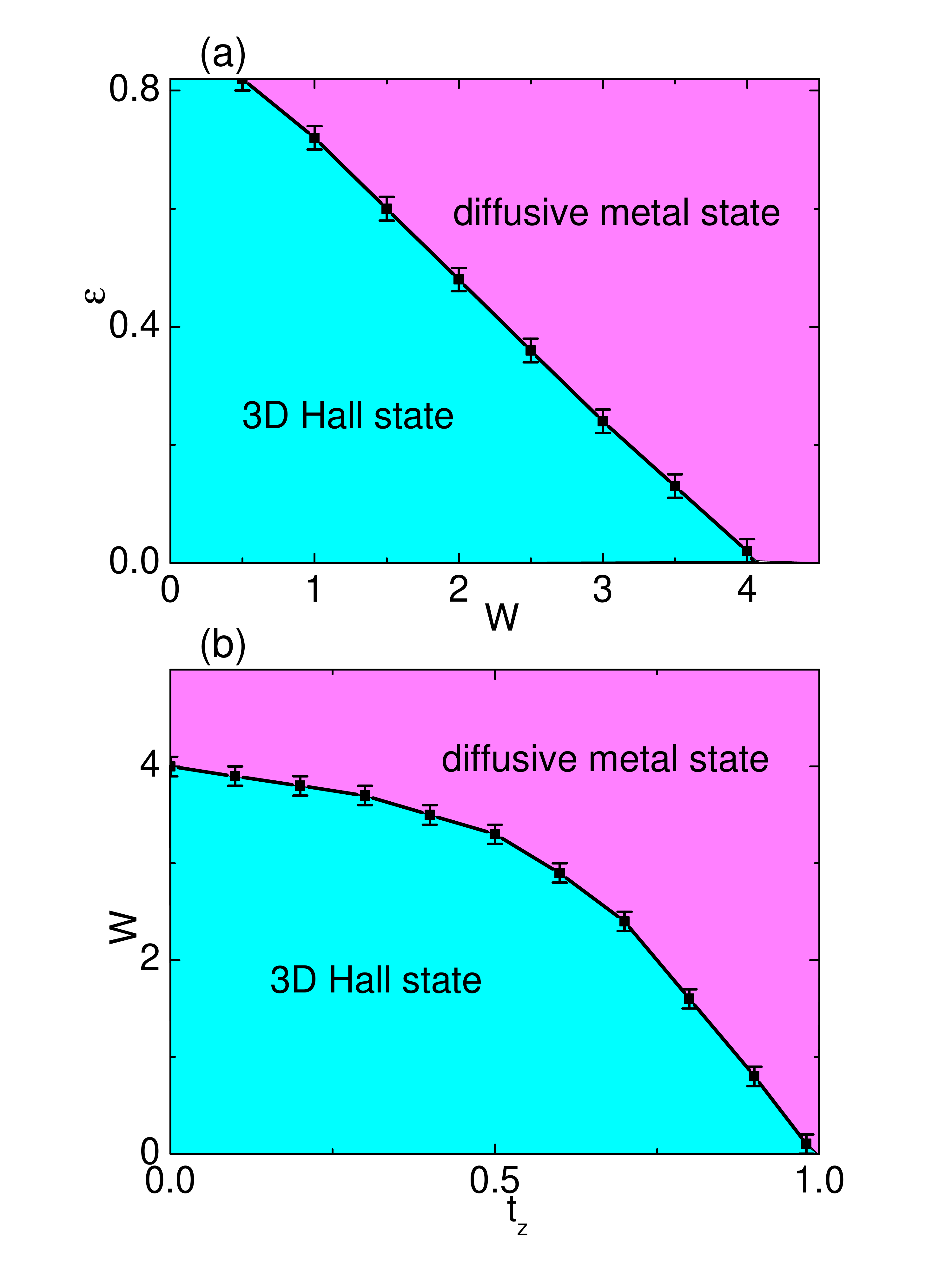}
	\caption{(Color online) The phase diagram separating the 3D Hall state from the diffusive metal state.  (a) is in the parametric space of $(W,\varepsilon)$ when $t_z=0$ and (b) is in $(t_z,W)$ when $\varepsilon=0$.  The uncertainty arises from determining the exact position of the phase transition.  The magnetic field is set as $B=\frac{2\pi}{60}$. }
	\label{Fig2}
\end{figure}

In Fig.~\ref{Fig1}, with the magnetic field $B=\frac{2\pi}{60}$, $\sigma_{xx}$ is plotted for different disorder strength $W$.  In Fig.~\ref{Fig1}(a), we consider the nontilted WSM of $t_z=0$.  Note that $\varepsilon_{1v}$ is indicated by the arrow.  In the quantum limit regime, when the Fermi energy $\varepsilon$ is around zero, $\sigma_{xx}$ is vanishingly small.  As $\varepsilon$ increases, $\sigma_{xx}$ remains unaffected.  Further increasing $\varepsilon$ to beyond the critical $\varepsilon_c$, $\sigma_{xx}$ becomes finite, meaning that the system at higher filling is driven into the diffusive metal state.  So we can use $\sigma_{xx}$ as an order parameter to determine the continuous phase transition from the 3D Hall state to the diffusive metal state.  The calculated phase diagram in the parametric space of $(W,\varepsilon)$ is plotted in Fig.~\ref{Fig2}(a).  It shows that $\varepsilon_c$ decreases linearly with $W$, suggesting that the increase of disorder can extend the diffusive metal state to the lower energy regime.  When $W\sim4.2$, $\varepsilon_c$ tends to be zero so that all states occupying the zeroth LL are driven into the diffusive metal state.  The phase diagram clearly reflects the competition between the Fermi energy and disorder in type-I WSMs.  The linear phase boundary can be qualitatively explained from the fact that the magnetoconductivity is closely related to the DOS $g(\varepsilon)$, $\sigma_{xx}(\varepsilon)\sim g(\varepsilon)$ [see Eq.~(11) below].  Indeed, as can be seen in Fig.~\ref{Fig1}(a), at weak disorder $W=0.2$, the DOS is less affected by disorder, so $\sigma_{xx}$ retrieves the sawtooth shape of the DOS in the clean type-I WSMs~\cite{J.Klier2015, Y.X.Wang2019}.  With increasing disorder, the $n\ge1$ Landau states are scattered into the quantum limit region.  In this case, the lowest scattered states that cause the transition into the diffusive metal state have to come from $n=1$ Landau states, and thus the movement of the states, which determines the critical $\varepsilon_c$, would be proportional to the disorder strength $W$.

As the disorder-induced scatterings play the decisive roles in forming the drift current along the electric field direction, $\sigma_{xx}$ increases with weak disorder.  This can be seen in the results of $W=0.5$ and $W=1$ in Fig.~\ref{Fig1}(a).  However, the oscillations in $\sigma_{xx}$ are suppressed by strong disorder $W\ge2$.  The reduction and eventual disappearance of the oscillations are ascribed to the change of the DOS, as it is smoothened by disorder.  In experiment, more oscillations in $\sigma_{xx}$ may be detected under the low magnetic field in a clean WSM sample~\cite{T.Liang, Y.Zhao}, as the energy separation between the neighboring LLs decreases for the low field. 

More importantly, in Fig.~\ref{Fig1}(a), we observe that $\sigma_{xx}$ is enhanced by disorder in the quantum limit regime, but is suppressed by strong disorder in the quantum oscillation regime.  So the effect of disorder on $\sigma_{xx}$ is distinct in the two regimes of type-I WSMs and could be understood with the help of the Einstein's relationship,  
\begin{align}
\sigma_{xx}(\varepsilon)=e^2Dg(\varepsilon), 
\end{align}
in which $D$ is the diffusion coefficient and $g(\varepsilon)$ is the DOS.  In the quantum limit regime, the disorder-induced scatterings dominate the magnetotransport along the external electric field.  Thus the diffusion coefficient $D$ gets enhanced with disorder and  $\sigma_{xx}$ increases.  While in the quantum oscillation regime, the disorder-induced LL broadening dominates the system, which in turn makes the DOS $g(\varepsilon)$ decrease with disorder~\cite{J.Klier2015} and correspondingly $\sigma_{xx}$ decreases. \textit{Such a behavior can be regarded as an important feature of disorder in 3D WSMs. }

The above effects of disorder still hold true when the Weyl cone is weakly tilted.  To see the influence of the Weyl cone tilting on the disorder-induced phase transition from the 3D Hall state to the diffusive metal state, we consider $\sigma_{xx}(0)$, the magnetoconductivity of the zero-energy state.  In Fig.~\ref{Fig2}(b), the calculated phase diagram is plotted in the parametric space of $(t_z,W)$.  It shows that the critical disorder $W_c$ decreases nonlinearly with $t_z$.  When $t_z\rightarrow1$, $W_c$ tends to zero.  This is because when the Weyl cone tilting increases, more electronic states may be scattered onto the zero energy state, making the 3D Hall state more susceptible to the diffusive metal state. 

For type-II WSMs, the results of $\sigma_{xx}$ are plotted in Fig.~\ref{Fig1}(b) of $t_z=1.5$.  As both the low-energy and high-energy LLs contribute to the magnetoconductivity, $\sigma_{xx}$ is pushed to a much large value of several tens of $\frac{e^2}{h}$.  So the type-II WSM can be regarded as a diffusive metal even in the clean case [see also the inset of Fig.~\ref{Fig3}(b)].  It is worth emphasizing that the high-energy LLs can be well described by the lattice model, but cannot be included in the low-energy $2\times2$ or $4\times4$ linearized model~\cite{V.A.Zyuzin, G.Sharma, K.Das}.  For weak disorder, $\sigma_{xx}$ increases with $W$ and is consistent with the above analysis in Fig.~\ref{Fig1}(a).  For strong disorder, $\sigma_{xx}$ decreases monotonously with $W$, as in type-II WSMs, the finite DOS around the band center decreases continuously by the disorder-induced scatterings~\cite{Y.X.Wang2019}.

\subsection{Magnetic Field}

\begin{figure}
	\includegraphics[width=8.2cm]{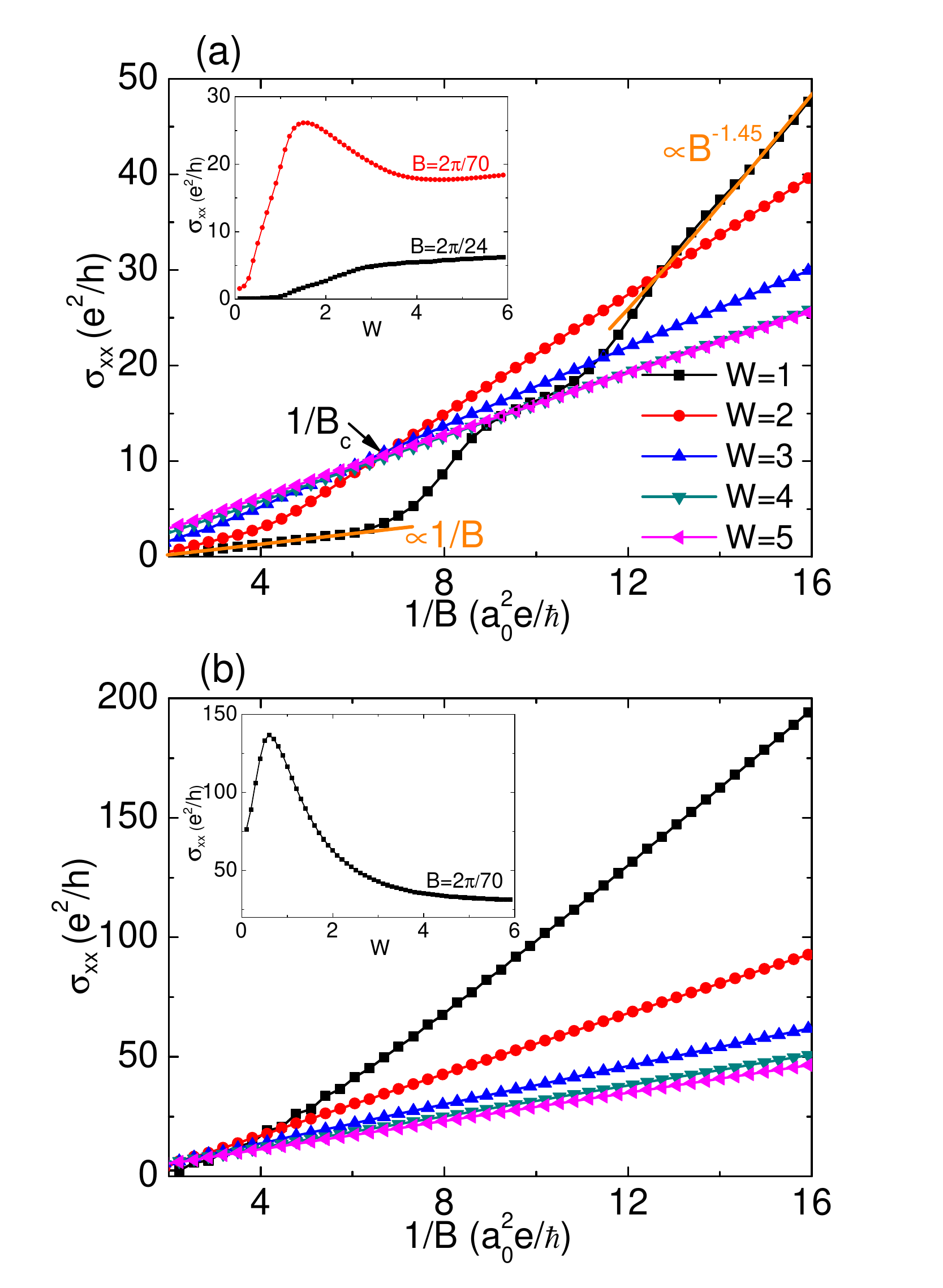}
	\caption{(Color online) The transverse magnetoconductivity $\sigma_{xx}$ vs the inverse magnetic field $\frac{1}{B}$ in tilted WSMs for different disorder strength $W$.  We set the parameters as $t_z=0.4$, $\varepsilon=1$ in (a) and $t_z=1.5$, $\varepsilon=0$ in (b).  In (a) of the curve $W=1$, the linear relationship $\sigma_{xx}\propto\frac{1}{B}$ is fitted in the high field regime and the power dependence of $\sigma_{xx}\propto B^{-1.45}$ is fitted in the low field regime.  The insets in (a) and (b) show the dependence of $\sigma_{xx}$ on $W$ for certain magnetic field $B$.  The legends are the same in both figures. }
	\label{Fig3}
\end{figure}

Next we investigate the influence of the magnetic field.  In Fig.~\ref{Fig3}, $\sigma_{xx}$ is plotted as a function of the inverse magnetic field $\frac{1}{B}$ for different disorder strength $W$.  Here we choose $\frac{1}{B}$ in order to better see the linear behavior of $\sigma_{xx}$.  In the numerical calculations, the minimum magnetic field can reach $B\sim\frac{2\pi}{100}$, as the low-field limit corresponds to excessively large magnetic supercell with large $L_y$, making the numerical calculations intractable.   

In type-I WSMs, from Eq.~(10), the vertex energy $\varepsilon_{1v}$ increases with the magnetic field.  Thus for the Fermi energy $0<\varepsilon<2(t-t_z)$, the relationship of $\varepsilon>\varepsilon_{1v}$ can be reversed as  $\varepsilon<\varepsilon_{1v}$.  This means that the magnetic field may drive the crossover of the system from the quantum oscillation regime into the quantum limit regime.  Consequently, according to the analysis in the previous section, the effect of disorder on $\sigma_{xx}$ may be changed.  This is demonstrated in Fig.~\ref{Fig3}(a) of $t_z=0.4$ and $\varepsilon=1$, where for the strong disorder $W\ge2$, when $B<B_c$ with $B_c\sim0.14$ being the critical value, $\sigma_{xx}$ decreases with $W$ and when $B>B_c$, $\sigma_{xx}$ increases with $W$. 

For all curves in Figs.~\ref{Fig3}(a) and (b), $\sigma_{xx}$ decreases with the magnetic field and is nonsaturated.  The decrease of $\sigma_{xx}$ is due to the deflection of the electron trajectories by the transverse Lorentz force.  The nonsaturated behavior is consistent with the experimental observations in 3D Dirac semimetal Cd$_3$As$_2$~\cite{T.Liang, Y.Zhao, J.Feng}.  For the weak disorder of $W=1$ in Fig.~\ref{Fig3}(a), in the high-field quantum limit regime, where only the zeroth LL contributes to the magnetoconductivity, $\sigma_{xx}$ is proportional to $\frac{1}{B}$, as can be seen by the fitted straight line.  In the middle-field regime, the nonmonotonic behavior of $\sigma_{xx}$ is exhibited, which is in line with the result in Ref.~\cite{X.Xiao}.  In the low-field regime, the power dependence of $\sigma_{xx}$ is shown and we fit the data as $\sigma_{xx}\propto B^{-1.45}$.  The ftted exponent of 1.45 is close to 1.67 found in Ref.~\cite{X.Xiao}. 
	 
At strong disorder, if the disorder-induced LL broadenings are larger than the separations between the neighboring LLs, the Landau quantizations are wiped out~\cite{J.Klier2015}.  Consequently, the significant impacts on the magnetoconductivity may be induced.  The numerical results in Fig.~\ref{Fig3}(a) show that at strong disorder, the excellent linear relationship of $\sigma_{xx}$ with $\frac{1}{B}$ is not limited in the quantum limit regime of type-I WSM, but can also be seen in the quantum oscillation regime.  In Fig.~\ref{Fig3}(b), the linear behavior in the whole magnetic field regime is also seen even in the overtilted $t_z=1.5$ WSMs.  From these results, we suggest that at strong disorder, the linear $\frac{1}{B}$ behavior of $\sigma_{xx}$ can be found in type-I and type-II WSMs for the uniformly distributed disorder, which broadens the understandings about the linear behavior of $\sigma_{xx}$. 

For the case of Coulomb disorder, the situation becomes a bit more complicated, but similar linear behavior was also obtained from the analytical calculations at zero temperature in nontilted WSMs when only the zeroth LL contributes to the magnetoconductivity~\cite{A.A.Abrikosov, X.Xiao, J.Klier2015}.  At nonzero temperature with Coulomb disorder, different regimes are revealed, depending on the relative magnitudes among the temperature $T$ and $\Omega$~\cite{J.Klier2015}, where $\Omega$ is the distance between the zeroth and $n=1$ LL.  The linear behavior still holds when $T$ is much smaller than $\Omega$~\cite{J.Klier2015}.  However, in other regimes, the linear behavior no longer persists~\cite{J.Klier2015}.  While in Ref.~\cite{X.Xiao}, it was reported that the linear magnetoconductivity is very robust against changes of temperature as long as the charge carriers come mainly from the zeroth Landau level.
 
In addition, in the insets of Fig.~\ref{Fig3}, $\sigma_{xx}$ is plotted as a function of $W$ for certain magnetic field.  It is interesting to find that in both type-I and type-II WSMs, $\sigma_{xx}$ gets saturated when disorder is strong enough.  The saturated behavior of $\sigma_{xx}$ may also be explained by the Einstein's relationship, where the strong disorder causes the dynamical equilibrium between the increasing diffusive coefficient and the decreasing DOS.  So the strong disorder can drive $\sigma_{xx}$ to reach its saturation value in WSM system, no matter what $t_z$ is.

\section{longitudinal magnetoconductivity}

\subsection{Chiral Anomaly} 

\begin{figure}
	\includegraphics[width=8.2cm]{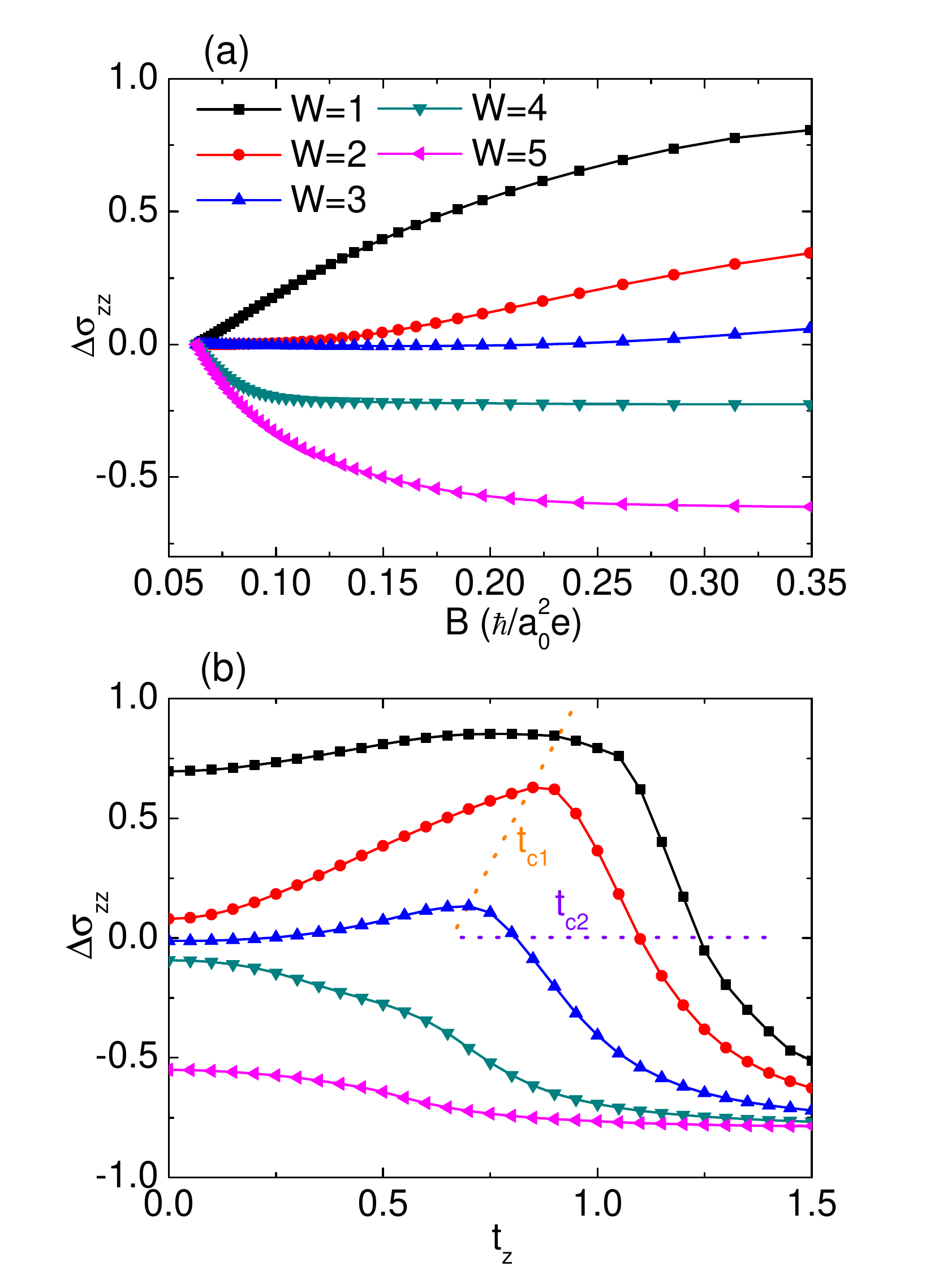}
	\caption{(Color online) The relative longitudinal conductivity $\Delta\sigma_{zz}$ in WSMs for different disorder $W$ with the Fermi energy set as $\varepsilon=0$.  (a) $\Delta\sigma_{zz}$ vs $B$ when $t_z=0.4$.  (b) $\Delta\sigma_{zz}$ vs $t_z$ when $B=\frac{2\pi}{20}$.  The critical values of $t_{c1}$ and $t_{c2}$ are indicated the dotted lines.  The legends are the same in both figures. }
	\label{Fig4}
\end{figure}

As the longitudinal magnetoconductivity $\sigma_{zz}$ occurs for the parallel electric field and magnetic field, $\boldsymbol E//\boldsymbol B$, the nonconservative electron density in different Weyl nodes and the chiral anomaly will be induced, which in turn changes the corresponding local Fermi energy in the Weyl node $\boldsymbol K_\eta$ as \cite{J.Behrends}
\begin{align}
\varepsilon_\eta=[\varepsilon^3+\frac{3}{2}\eta \hbar (v^2-v_z^2)^{\frac{3}{2}}e^2\tau_v \boldsymbol E\cdot\boldsymbol B]^{\frac{1}{3}}, 
\end{align}
where $\tau_v$ denotes the internode relaxation time.  Note that the Weyl cone tilting has been included in the above equation.   

\begin{figure}
	\includegraphics[width=8.8cm]{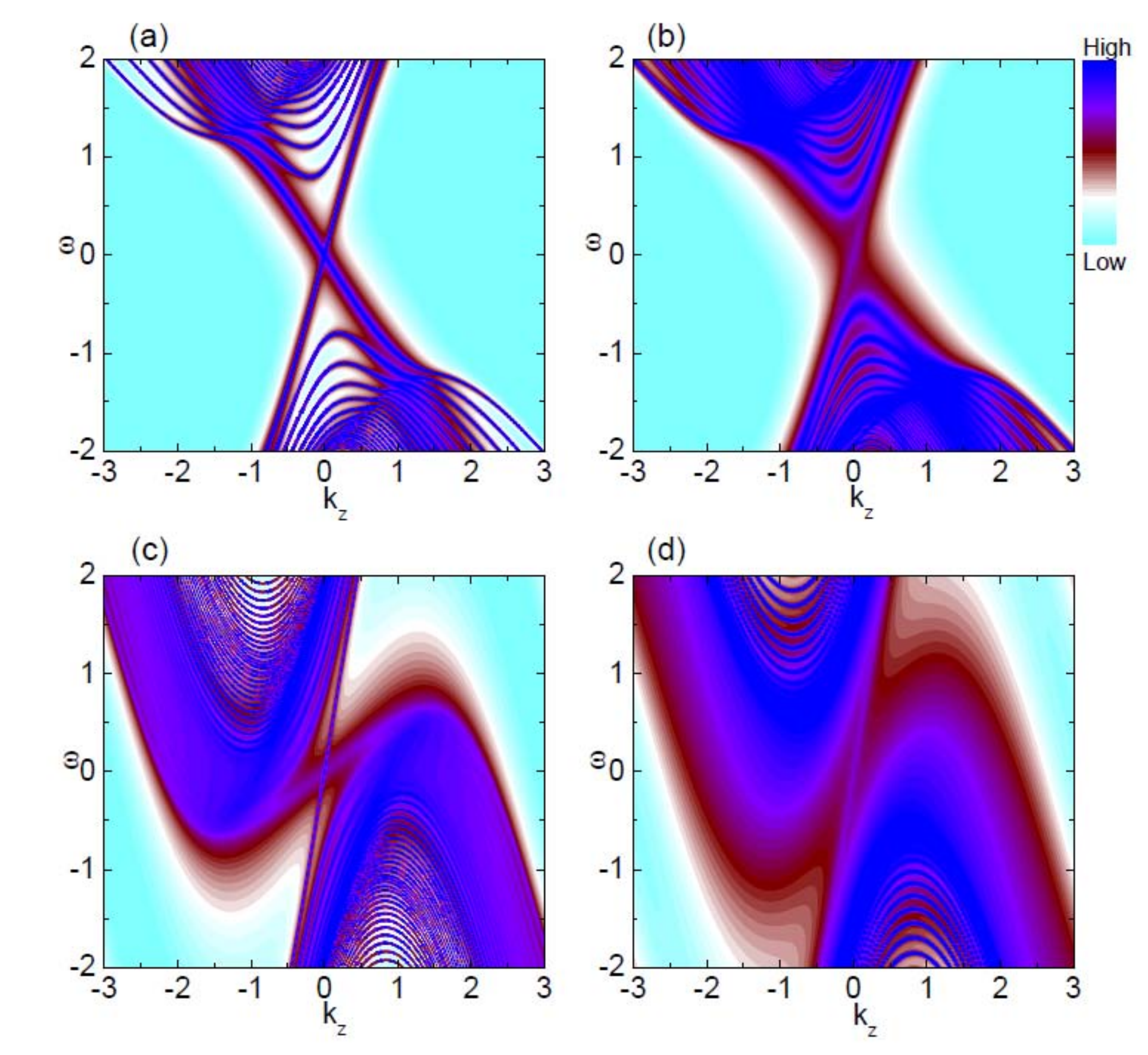}
	\caption{(Color online) Contour plot of the spectral function $A(k_z,\omega)$ for the disordered WSMs when $B=\frac{2\pi}{60}$.  The Weyl cone tilting is $t_z=0.4$ in (a) and (b), $t_z=1.3$ in (c) and (d).  The disorder strength is $W=1$ in (a) and (c), $W=4$ in (b) and (d).}
	\label{Fig5}
\end{figure}

We consider the relative longitudinal conductivity of $\Delta\sigma_{zz}=\frac{\sigma_{zz}-\sigma_m}{\sigma_m}$, where $\sigma_m=\sigma_{zz}(B=\frac{2\pi}{100})$ is the conductivity of the minimum magnetic field that we can reach in the numerical calculations.  Since the magnetic field is small enough, using $\sigma_m$ as the reference to calculate $\Delta\sigma_{zz}$ would give the results that are qualitatively valid.  With $t_z=0.4$ and $\varepsilon=0$, $\Delta\sigma_{zz}$ is plotted as a function of the magnetic field $B$ in Fig.~\ref{Fig4}(a) for different disorder strength $W$.  For weak disorder $W\le2$, the positive magnetoconductivity is clearly observed in $\Delta\sigma_{zz}$.  However, the positive magnetoconductivity is suppressed at $W=3$, and even turns to be negative at strong disorder $W\ge4$.  It also shows that at high magnetic field, $\Delta\sigma_{zz}$ tends to be saturated.  This observation is consistent with the previous results of the $B-$independent conductivity in the strong field limit~\cite{H.Z.Lu, V.Aji}, but is quite different from the nonsaturated behavior of $\sigma_{xx}$ studied in the previous section.  

To see the effect of the Weyl cone tilting $t_z$ on $\Delta\sigma_{zz}$, in Fig.~\ref{Fig4}(b), we plot $\Delta\sigma_{zz}$ as a function of $t_z$ at the fixed magnetic field $B=\frac{2\pi}{20}$.  For weak disorder $W=1$, $\Delta\sigma_{zz}$ is positive at $t_z=0$ and then increases with $t_z$.  When $t_z$ crosses the first critical point $t_{c1}$, $\Delta\sigma_{zz}$ gradually decreases and when $t_z$ crosses the second critical point $t_{c2}$, $\Delta\sigma_{zz}$ becomes negative.  Note that the transition of $\Delta\sigma_{zz}$ from positive to negative is continuous and does not happen at the Lifshitz transition point of $t_z=1$, where the WSM system changes from type-I to type-II.  So there are no qualitative changes of $\Delta\sigma_{zz}$ around $t_z=1$, which is in line with the semiclassical study of $\frac{\sigma_{xx}}{\sigma_{zz}}$ and log$(\frac{\sigma-\sigma_{B=0}}{\sigma_{B=0}})$ in tilted WSMs~\cite{G.Sharma}.  With the increase of $W$, both the critical $t_{c1}$ and $t_{c2}$ move to the weaker values, as shown by the dotted lines in Fig.~\ref{Fig4}(b).  For the strong disorder $W\ge4$, the negative $\Delta\sigma_{zz}$ is clearly seen for all $t_z$. 

In Fig.~\ref{Fig4}(b), the observed negative $\Delta\sigma_{zz}$ at large $t_z$ and weak $W$ is supported by another work based also on the lattice model~\cite{Y.W.Wei}, but is in sharp contrast with the positive $\Delta\sigma_{zz}$ in the previous work based on the low-energy model~\cite{V.A.Zyuzin, G.Sharma, K.Das}.  We may also attribute the negative $\Delta\sigma_{zz}$ to the reason that the lattice model incorporates the contributions to the magnetoconductivity from the low-energy Landau states as well as the high-energy ones.  According to this, an important question that whether the chiral anomaly in tilted WSMs is preserved or not at strong disorder cannot be simply judged from the sign of $\Delta\sigma_{zz}$.  
 
To investigate the above question, we use the criteria that the chiral anomaly could still be present as long as the Weyl nodes remain gapless~\cite{A.A.Burkov2015, H.Z.Lu, H.W.Wang, E.V.Gorbar}.  If the Weyl nodes are gapless, the chiral symmetry of the Weyl fermions is preserved.  While if the Weyl nodes are gapped, the Weyl fermions acquire mass and the chiral symmetry is broken.  Here we try to make judgments by calculating the spectral function $A(k_z,\omega)$ from the disorder-averaged Green's function $\bar G^R$ in Eq.~(7)~\cite{S.Datta}, 
\begin{align}
A(k_z,\omega)=-\frac{1}{\pi}\text{Im} \bar G^R(k_z,\omega). 
\end{align} 

The contour plots of $A(k_z,\omega)$ for the disordered WSMs are given in Fig.~\ref{Fig5}.  For $t_z=0.4$, in Fig.~\ref{Fig5}(a) of weak disorder $W=1$, the Weyl nodes can still be seen, although the spectrum broadening is present.  In Fig.~\ref{Fig5}(b) of strong disorder $W=4$, besides the further spectrum broadening, the low-energy $n\ge1$ LLs move to the band center.  For the chiral zeroth LLs, the $\eta=-1$ branch is still distinguishable, but another $\eta=1$ branch is blurred by strong disorder, as the DOS of $\eta=-1$ branch is larger than that of $\eta=1$ in the clean case~\cite{Y.X.Wang2019},
\begin{align} 
g_{n=0,\eta=-1}(\varepsilon)\propto
\frac{1}{|v-v_z|}>g_{n=0,\eta=1}(\varepsilon)\propto\frac{1}{v+v_z}.  
\end{align}
Evidently, as disorder increases, more states are scattered to be around zero energy, driving the system into the diffusive metal state.  This is consistent with the previous analysis about $\sigma_{xx}$.  Similar conclusions can also be derived from Figs.~\ref{Fig5}(c) and (d) for the overtilted case $t_z=1.3$, although the Weyl nodes are concealed in the high-energy states.  As there is no gap opening around the Weyl nodes, we come to the conclusion that the chiral symmetry is preserved in the tilted WSMs and thus the chiral anomaly will not be broken by strong disorder.

\subsection{Shubnikov de-Haas Oscillations}

\begin{figure}
	\includegraphics[width=8.2cm]{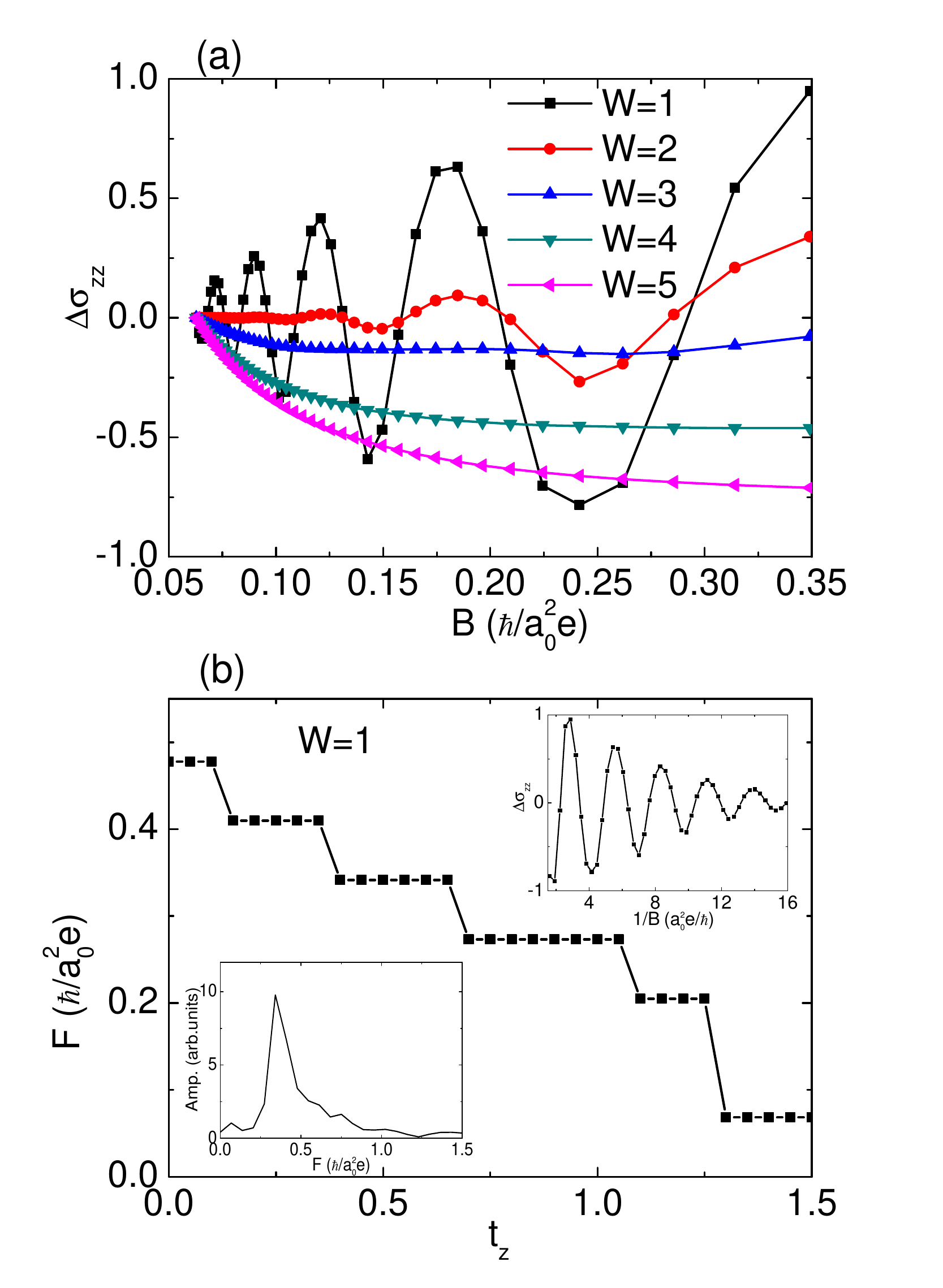}
	\caption{(Color online) The SdH oscillations in the relative longitudinal conductivity $\Delta\sigma_{zz}$ in WSMs, with the Fermi energy set as $\varepsilon=0.2$.  (a) $\Delta\sigma_{zz}$ vs the magnetic field for different disorder $W$ for the tilting parameter $t_z=0.4$.  (b) The frequency of the oscillations vs $t_z$ for $W=1$.  The upper inset gives $\Delta\sigma_{zz}$ vs $\frac{1}{B}$ while the lower inset is the FFT analysis with one peak frequency for $t_z=0.4$. }
	\label{Fig6}
\end{figure}

It is worthy of noting that in the previous work~\cite{Y.W.Wei}, the SdH oscillations in the longitudinal magnetoconductivity were exhibited at zero Fermi energy and can be attributed to the complicated Fermi surface in their model.  While in our model, when the Fermi energy is zero, as in Eq.~(12), the chiral anomaly will cause the local Fermi energies in the two Weyl nodes symmetric to the zero energy, $\varepsilon_+=-\varepsilon_-$.  Thus when the magnetic field changes, no net LLs cross the Fermi surface and no oscillations are observed in $\Delta\sigma_{zz}$, as in Fig.~\ref{Fig4}(a).  

To see the SdH oscillations, a necessary condition is the asymmetric local Fermi energies, $\varepsilon_+\neq\varepsilon_-$, which requires a nonvanishing Fermi energy, $\varepsilon\neq0$.  Here we set $\varepsilon=0.2$ and plot the relative $\Delta\sigma_{zz}$ as a function of the magnetic field $B$ in Fig.~\ref{Fig6}.  In Fig.~\ref{Fig6}(a), it shows that at weak disorder $W\le2$, the oscillations are evidently present, with the amplitude growing with magnetic field.  When disorder increases, the oscillations are suppressed, due to the reduction of the relaxation time~\cite{Y.W.Wei}.  At strong disorder $W\ge4$, the oscillations disappear completely and the results are similar to Fig.~\ref{Fig4}(a).  Therefore, to observe the SdH oscillations in $\sigma_{zz}$ experimentally, a clean WSM sample is needed, which is the same requirement as that in $\sigma_{xx}$.

We make further analysis of the SdH oscillations in $\Delta\sigma_{zz}$ at weak disorder of $W=1$.  According to the Lifshitz-Kosevich formula~\cite{D.Shoenberg, Y.Zhao}, the relative magnetoconductivity is periodic with the inverse magnetic field $\frac{1}{B}$ and a sinusoidal relation is given by 
\begin{align}
\Delta\sigma_{zz}\propto\text{cos}[2\pi(\frac{F}{B}+\varphi)],
\end{align}
where $F$ is the frequency and $\varphi$ is the phase shift.  Note that the frequency will not be changed by the choice of $\sigma_m$ at the nonvanishing magnetic field.  In Fig.~\ref{Fig6}(b), we plot the frequency $F$ extracted from the fast Fourier transformation (FFT) as a function of the Weyl cone tilting $t_z$.  To extract $F$, we take $t_z=0.4$ as an example.  In the upper inset of Fig.~\ref{Fig6}(b), $\Delta\sigma_{zz}$ is plotted vs the inverse magnetic field $\frac{1}{B}$ and then in the lower inset of Fig.~\ref{Fig6}(b), the FFT results are given.  From FFT, the single fundamental frequency for $t_z=0.4$ is extracted directly as $F=0.41\frac{\hbar}{a_0^2e}$.  The occurrence of the single oscillation frequency is due to the simple Fermi surface in our model.  If the Fermi surface is complicated, the number of the oscillation frequency may be two~\cite{Y.Zhao} or even more~\cite{Y.W.Wei}.  

In Fig.~\ref{Fig6}(b), we observe that the frequency $F$ exhibits discrete steps with the Weyl cone tilting $t_z$ and decreases gradually.  The behavior of $F$ can be understood as follows.  When $t_z$ increases, the LLs, together with the local Fermi energy will move to the zero energy [see Eqs.~(10) and~(12)].  The appearance of the frequency step is because the minor change of $t_z$ cannot cause the LLs cross the local Fermi energy.  But when $t_z$ increases a lot, if the decrease of the Fermi energy surpasses the moving of LLs, $\Delta\varepsilon_\eta>\Delta\varepsilon_{nv}$, less LLs are driven to cross the local Fermi energy by changing the magnetic field and therefore the oscillation frequency $F$ decreases.  The behavior of $F$ is similar when $t_z>1$,  meaning that the SdH oscillations are well kept in type-II WSMs.  So for the SdH oscillations, there is again no qualitative difference between type-I and type-II WSMs.

\section{Discussions and Conclusions}

Compared with the uniformly distributed disorder model, there is another important disorder model of the Gaussian distribution.  As the two disorder models share certain similarities in determining the physical properties of the 3D Dirac/Weyl semimetal system, the uniformly distributed disorder model used here is reliable.  For example, in Ref.~\cite{J.H.Pixley} of a Dirac semimetal system, it is found that the critical exponent $z$, characterizing the correlation in time, has the same value in the two disorder models.  In another work of a WSM systems~\cite{S.Bera}, the authors do the numerical calculations with the uniformly distributed disorder model and make the renormalization group (RG) analysis with the Gaussian distribution.  They suggested that the critical exponent $z$ obtained from the RG analysis is in agreement with the numerical findings.  

For the real WSM material WTe$_2$~\cite{C.Wang}, the hopping integral is taken as $t=0.1$eV and the lattice constant $a_0=10$\AA.  Then we estimate that the disorder strength $W=1$ to be $\sim0.1$eV, and the magnetic field unit $\frac{\hbar}{ea_0^2}\sim 656$T.  In our numerical calculation, the smallest magnetic field is 0.0628 unit, corresponding to real value of 40T, and we use up to 200T in the calculations.  Such colossal fields in condensed matter systems should allow for many interaction effects enhanced or induced by magnetic field.  For example, the gap would be generated via the magnetic catalysis effect~\cite{V.A.Miransky}.  Here we do not consider such effects, but only use very large magnetic fields to simplify the numerical calculations and work in the regime where only a few LLs are filled. 

We have so far ignored the Zeeman term due to the magnetic field.  Its effect is to split the spinless bands into $\varepsilon_\uparrow=\varepsilon+\frac{1}{2}g\mu_BB$ and $\varepsilon_\downarrow=\varepsilon-\frac{1}{2}g\mu_BB$~\cite{A.A.Burkov2015}, with $g$ being the Lande factor and $\mu_B$ the Bohr magneton.  If $\varepsilon_{1v}-\frac{1}{2}g\mu_BB>0$, the quantum limit regime is still present in WSM1, with the critical line driven by disorder in Fig.~\ref{Fig2}(a) moving downward as $\varepsilon_c\rightarrow\varepsilon_c-\frac{1}{2}g\mu_BB$.  While if $\varepsilon_{1v}-\frac{1}{2}g\mu_BB<0$, the vertex energy of $n=1$ LL crosses the zero energy.  In this case, the system may behave a bit similar to a type-II WSM.  The other conclusions of the magnetoconductivity obtained in this work, such as the linear relationship with $\frac{1}{B}$, the chiral anomaly and the SdH oscillations, are under the condition of strong disorder and keep unchanged with the varying Fermi energy.  We speculate that these conclusions will not be affected by the Zeeman effect. 

Finally, we make some comparisons with the effect of disorder on the Hall conductivity $\sigma_{xy}$ in WSMs.  (i) $\sigma_{xy}$ can be nonzero even in the clean case~\cite{A.A.Burkov2011}, as it is due to the electron moving around a circle by the Lorentz force.  While for the diagonal magnetoconductivity $\sigma_{xx}$ and $\sigma_{zz}$, disorder is an indispensable factor in forming the magnetotransport.  (ii) At weak disorder, $\sigma_{xy}$ exhibits certain robustness in type-I WSMs and the robustness is broken successively from the higher LLs to the lower ones~\cite{Y.X.Wang2019}.  For $\sigma_{xx}$ and $\sigma_{zz}$, the robustness to disorder is absent.  (iii) At strong disorder, $\sigma_{xy}$ is completely suppressed in both type-I and type-II WSMs, as the Hall states carrying opposite Chern numbers are annihilated with each other~\cite{Y.X.Wang2019}.  For $\sigma_{xx}$ and $\sigma_{zz}$, their magnitudes cannot be suppressed completely, due to the different mechanism caused by disorder.  Instead, the numerical results suggest that strong disorder can drive $\sigma_{xx}$ to reach its saturation value, and $\sigma_{zz}$ to reach a rather large negative value.

To summarize, in this work, based on the minimum lattice model and quantized LLs, we have studied the effects of disorder on the transverse and longitudinal magnetoconductivity in tilted WSMs.  As the lattice model correctly describes the low-energy LLs as well as the high-energy ones, it goes beyond the semiclassical theory and can capture the main physics related to the magnetotransport.  We find that there exists evident difference of $\sigma_{xx}$ with disorder between type-I and type-II WSMs.  While for the linear behavior in $\sigma_{xx}$, the chiral anomaly and the SdH oscillations in $\sigma_{zz}$, there are no evident differences between type-I and type-II WSMs.  Although only the WSM model of the minimum number of nodes is considered, the obtained results are quite reliable as long as the bulk physics is focused on.  The effect of multiple nodes can be accounted for by simply multiplying the results of the single-node model by the number of the pairs of nodes.  There are also several open questions of the magnetotransport in WSMs that are left for the future works, such as the effect of the mass term~\cite{H.W.Wang, V.Konye} and the chiral anomaly at strong disorder.  More theoretical and experimental works about the magnetotransport in WSM systems are expected in the future.

\section{Acknowledgments}

This work was supported by NSFC (Grants No. 11804122 and No. 11905054), and the Fundamental Research Funds for the Central Universities of China.

\end{document}